# The new FAST module: a portable and transparent add-on module for time-resolved investigations with commercial scanning probe microscopes

Carlo Dri [a,b], Mirco Panighel [b,*], Daniel Tiemann [c], Laerte L. Patera [b,d,1], Giulia Troiano [d,e,2], Yves Fukamori [e], Fabian Knoller [e], Barbara A. J. Lechner [e], Giuseppe Cautero [a], Dario Giuressi [a], Giovanni Comelli [b,d], Jordi Fraxedas [c], Cristina Africh [b], Friedrich Esch [e]

[a] Elettra-Sincrotrone Trieste, S.S. 14 km 163.5, Basovizza, I-34149 Trieste, Italy.
[b] CNR-IOM Laboratorio TASC, S.S. 14 km 163.5, Basovizza, I-34149 Trieste, Italy.
[c] Catalan Institute of Nanoscience and Nanotechnology (ICN2), CSIC and BIST, Campus UAB, Bellaterra, 08193 Barcelona, Spain.
[d] Department of Physics, University of Trieste, via A. Valerio 2, I-34127 Trieste, Italy.
[e] Chair of Physical Chemistry, Department of Chemistry & Catalysis Research Center, Technical University of Munich, Lichtenbergstr. 4, 85748 Garching, Germany.

[1] Present address: Institute of Experimental and Applied Physics, University of Regensburg, D-93053 Regensburg, Germany.
[2] Present address: Telit Communications Spa, Via Stazione di Prosecco 5b, Sgonico I-34010 Trieste, Italy.

* Corresponding author.
E-mail address: panighel@iom.cnr.it

## Abstract

Time resolution is one of the most severe limitations of scanning probe microscopies (SPMs), since the typical image acquisition times are in the order of several seconds or even few minutes. As a consequence, the characterization of dynamical processes occurring at surfaces (e.g. surface diffusion, film growth, self-assembly and chemical reactions) cannot be thoroughly addressed by conventional SPMs. To overcome this limitation, several years ago we developed a first prototype of the FAST module, an add-on instrument capable of driving a commercial scanning tunneling microscope (STM) at and beyond video rate frequencies. Here we report on a fully redesigned version of the FAST module, featuring improved performance and user experience, which can be used both with STMs and atomic force microscopes (AFMs), and offers additional capabilities such as an atom tracking mode. All the new features of the FAST module, including portability between different commercial instruments, are described in detail and practically demonstrated.

## 1. Introduction
The invention of scanning probe microscopes in the 1980s has opened the unprecedented possibility of imaging and controlling matter at the molecular and atomic scale. The publishing rate of research articles where SPMs are adopted has grown steadily through the years, exceeding 7 k/year in 2017 [1]. The versatility of these instruments is one of their main strengths: they can be operated in a multitude of environments, from vacuum, to air and high pressure cells, in liquids or electrochemical cells.

However, one of their fundamental limitations is the time resolution, since the majority of commercially available instruments offers typical acquisition times per image of seconds or even minutes, depending on imaging conditions. This hinders the observation of many dynamical phenomena of fundamental and technological interest that commonly occur on much faster time scales and, in the case of relatively large activation energies, cannot be slowed down by lowering the experiment temperature. In the past, several attempts have been made in order to achieve high scanning frequencies, typically by designing dedicated instruments with high resonance frequency scanners [2-4], in some cases even performing the slow and fast movements by separate scan heads [5, 6] or different scanner shapes and probe trajectories [7]. In combination with fast feedback electronics and preamplifiers, frame rates up to 200 images/s could be reached in a hybrid mode between the constant-height and constant-current [4, 8]. Overall, several successful solutions have been proposed with dedicated systems for FastSTM [2-9], FastAFM [10-15] and atom tracking [16, 17].

In contrast to these approaches, we accelerate existing, slow instruments without modifying the existing scanner heads, exploiting the fact that the majority of the mechanical problems encountered in fast, constant-current measurements can be avoided when measuring in a quasi-constant-height mode [2]. Based on this concept, back in 2011, following several years of prototyping and experiments, we developed a unique modular solution and introduced the first prototype of the FAST module [18, 19], an add-on electronics applicable in principle to any existing scanning probe microscope. The basic concept behind the module was to simply "piggyback" the existing electronics of the commercial SPM in a completely transparent fashion, by generating the fast scan signals and adding them to the original, slow scan signals. The probe signal was then acquired independently by the module itself. In this way, the module could be installed without modification of the existing scanner hardware and electronics and enabled seamless switching between the fast and slow scan modes.

Against the common belief that mechanical resonances of the SPM scanners pose an insurmountable limitation to the scan speed, with our initial prototype, we successfully demonstrated that imaging at high speed with fast scan frequencies among and above such resonances is indeed possible with a resolution comparable to the slow scan mode [19, 20].

## 2. Limitations of the first prototype

The first prototype of the FAST module, described in detail in reference [18], suffered from a number of limitations. First of all, the whole module was based on an embedded personal computer (PC) (PC/104 architecture) hosting one Field Programmable Gate Array (FPGA) and custom made (digital to analog converter) DAC and (analog to digital converter) ADC boards, mounted on two Industry Standard Architecture (ISA) bus slots. The nominal data transfer rate of ISA bus, about 8 MB/s, together with 10 Mbit/s of the Ethernet connection imposed two major bottlenecks to the data transfer rates; moreover, while the generation of the scanning signals was carried out by the FPGA controlling the DACs, the data acquisition chain was affected by unpredictable interrupts from the operating system and by the small size of the on-board memories, leading to random loss of data points.

The 8 bit resolution of the DACs was also a strong limitation to the achievable imaging quality: the harmonic distortion of the generated fast sine waves was large enough to cause instabilities and distortions in the images, since clear improvements could be seen by "cleaning" the scan waves with strong band-pass filtering. On the ADC side, the 12 bit, 10 MS/s sampling was also a limiting factor: in fact, given that fast imaging requires pixel acquisition frequencies on the order of hundreds of kHz, the bandwidth of commercial preamplifiers (e.g. about 200 kHz with $10^8$ V/A gain) is well below this requirement, and thereby strong oversampling is required to recover a useful signal when working above the preamplifier cutoff.

Moreover, the sample tilt correction in the direction orthogonal to the sample surface was carried out manually with a custom-built analog box, where correction parameters were tuned by means of analog potentiometers, making it difficult to re-use parameter sets that were known to be working.

We should also mention that a proper, open and extensible data analysis framework was also missing, making it difficult to implement custom data exploration tasks.
Finally, although the portability of the FAST module between different STMs and its extension to AFMs was proposed as a possible development, its practical realization was never performed.

## 3. Design and specifications of the new FAST module

The new FAST module, albeit based on the exact same concept, is completely new from the hardware, firmware and software perspective, implementing in a reliable and user-friendly way many of the technical solutions adopted in the previous prototype. The module is composed of an acquisition and control electronics and a high-voltage summation device (see Figure 1). The control unit consists of a commercial PXI system by National Instruments hosting an FPGA-based high-speed reconfigurable input/output board (model NI PXI-7951). A NI-5781 adapter module provides two DACs for the generation of the fast scan waveforms (with an output range of 2 V peak to peak) and two ADCs for the acquisition of the fast measurement signals (with an input range of 1 V peak to peak) driven by the same clock as the DAC, thereby ensuring stable and reliable synchronization between probe motion and data acquisition. It is worth noting that the high-speed (16 bit @ 100 MS/s) of the DACs used to control the probe movement is crucial to avoid any scan distortion due to resonances caused by overtones. While the typical bandwidth required for fast SPM measurements stays in the MHz frequency range, commercially available preamplifiers provide a rather smaller bandwidth (typically around 45 kHz in the $10^9$ V/A range and up to 200 kHz in the $10^8$ V/A range), thus leading to a strong suppression of the signal in the frequency range of interest. In this context, the high sampling rate (14 bit @ 100 MS/s) of the FAST module ADCs allows oversampling the input in order to recover the measured signal even well above the preamplifier cutoff. When taking fast images with pixel frequencies above the preamplifier cutoff frequency, the preamplifier will increasingly act as an integrator. This effect can partially be removed *a posteriori* by the use of a lateral gradient filter in order to recover images with standard constant-height contrast. When studying surface dynamics, such as diffusional jumps or molecule rotations, the accessible time resolution is, however, much lower than the pixel frequency: if the statistical analysis strictly requires the detection of single events, moving features have to be detected, on average, at least in two subsequent images; the achievable resolution is thus less or equal half the image frequency.

A slower, multi-functional input/output board (model NI PXIe-6341 or NI PXI-6221) is used for the generation of the slow scan signals and the acquisition of other useful physical quantities, such as sample temperature or tunneling bias. The generated scan voltages are then routed into a new high voltage summation device based on APEX PA90 amplifiers with a ±160 V output range and up to 20 kHz bandwidth and then added to the corresponding signals from the SPM control unit. In addition to the *x* and *y* summation channels, a new channel is dedicated to the *z* direction and the scan voltages can be optionally amplified (up to 4.8 times) to reach the desired scan range. The output of the summation device is then sent to the piezoelectric motor of the SPM head. The response signals from the SPM are finally acquired by the control electronics of the FAST module. A passive attenuator is optionally used to reduce the typical ±10 V output signal range of the SPM control unit to comply with the 1 V peak to peak input of the ADCs.

A high-level LabVIEW software running on a host PC drives the control electronics and provides a dedicated graphical user interface (GUI) for configuring the scan parameters and visualizing and storing the acquired data. To limit the need of resources on the host PC, rather than displaying the movie in real time, a selected frame rate (usually few frames per second) is set for data visualization. A high-efficiency integrated software suite is needed to account for real time processing and storing of the data acquired with the FAST module. As an example, the stream of data during the acquisition of a 200 × 200 pixels movie at an image rate of 100 frames per seconds with a resolution of 14 bit amounts to 56 Mbit/s. In order to conveniently and efficiently store this data, movies are saved into hierarchical data format (HDF5) together with their relevant metadata. A dedicated open source Python-based software (see section 8) was developed to visualize, process, analyze and export the movies.

The control of the FAST module is entirely provided through an integrated hardware/software framework allowing the reproducibility and stability required for reliable data acquisition to be achieved. In particular, the adoption of an FPGA-based architecture guarantees a reproducible and stable delay between probe movement and data acquisition. At the same time, the availability of both slow and fast inputs gives the full freedom to add fast dithering/scanning signals and to track slow drift issues.

As can be seen from the schematics of Figure 1, the new FAST module, as its predecessor, is completely transparent with respect to the existing electronics: when the output from the FAST control unit is set to zero, the SPM can be driven by the conventional acquisition software. At the same time its design allows a seamless switch between the fast mode and the traditional, slow imaging mode.

In the following sections, the functionality of the FAST module will be described in detail for its different operation modes, namely FastSTM, FastAFM and atom tracking.

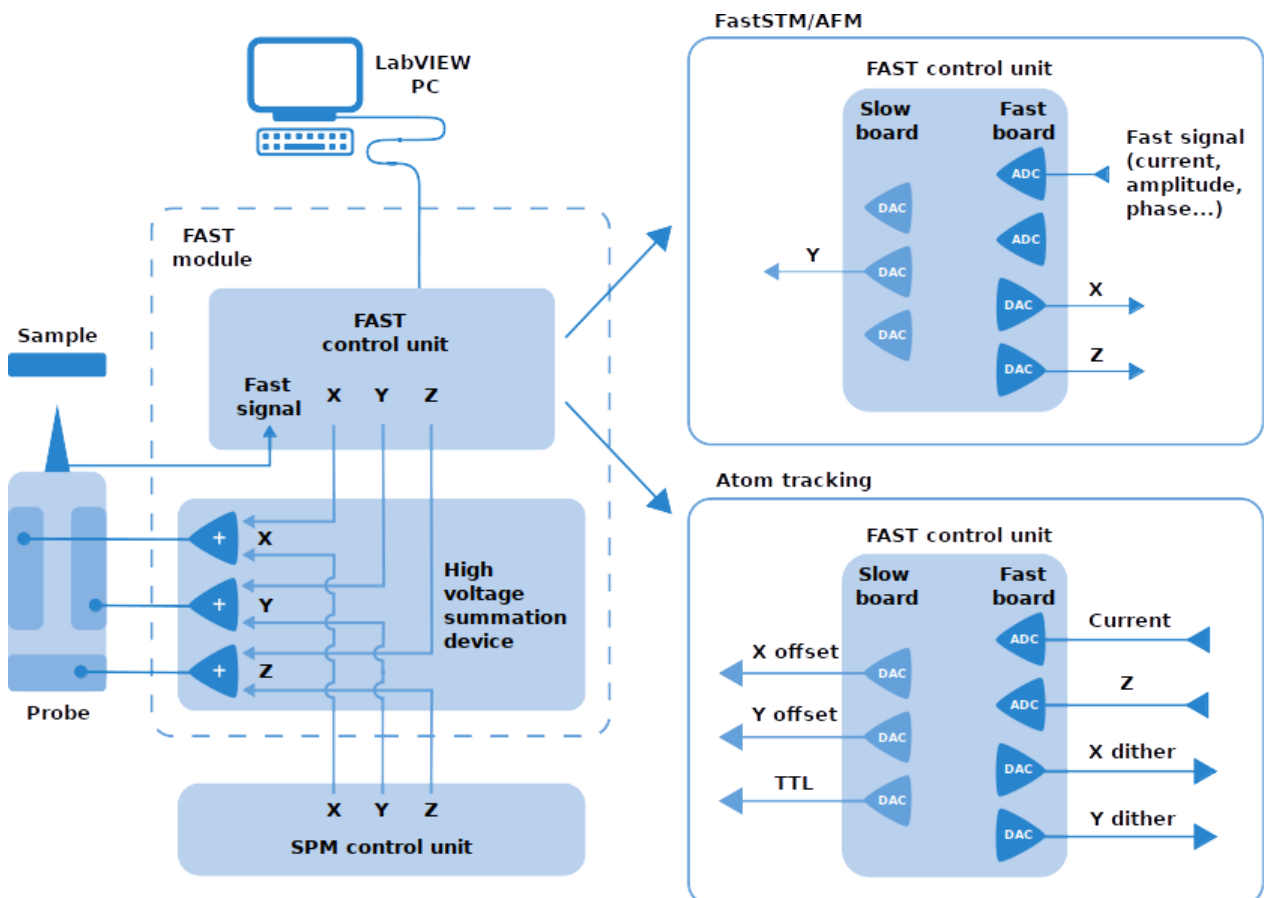

Figure 1 - Schematics of the FAST module working principle (left). Connection scheme of the FAST module for the FastSTM/AFM and atom tracking operation modes (right).

## 4. FastSTM

In the FastSTM mode, the high-speed DACs are used to output the fast $x$ and $z$ scan waveforms, while the slow $y$ scan signal is generated by the conventional DAC board. As already pointed out previously [18], the generated scan waveform for the fast signals, whose frequency can reach up to 20 kHz, has to describe a sine function as accurately as possible in order not to excite uncontrolled resonances of the scanner. Furthermore, to avoid inducing transient responses due to abrupt changes in the probe movement, the scan amplitude is smoothly ramped up to the setpoint at the beginning, and down to zero at the end of the probe scan. For the slow scan signal we instead use a simple triangular waveform, since its frequency during the use of the FAST module usually does not exceed 50 Hz.

When scanning at a speed much higher than the one for which the scanner has been designed, the phase of the probe's mechanical response changes with respect to the phase of the excitation signal, due to the combined effects of an increased hysteresis of the piezoelectric scanner and the approaching of its resonance regimes. Moreover, the different paths of the signals inside the electronics can introduce additional delays. To take these effects into account, the FAST module is provided with a tunable phase shift between the signal acquisition and the fast $x$ signal generation, which is taken as a reference.

The fast $x$ and $z$ signals are identically generated, the latter with a variable amplification and an additional phase shift. In this case, the control of the $z$ signal amplification is used to provide a slope compensation to overcome the unavoidable sample tilting, while the phase shift, as pointed out for the signal acquisition, is required to precisely match the actual movement along the $x$ direction. In this new version of the module, the phase shift and amplitude of the $z$ signal - which were previously manually controlled through an external analog module - are integrated into the control software, allowing the user to easily re-use working parameters sets and thereby improving the reproducibility of the measurements.

The frequency response of conventional feedback control electronics lies well below 20 kHz. For this reason, scans with the FAST module are performed in quasi-constant height mode, *i.e.* a relatively low value of the feedback loop is set. In this condition, along the fast scan direction, the probe moves at a constant height (besides the additional slope correction), the feedback acting as an integrator, while along the slow scan direction the feedback accounts for any sample tilting and thermal drift. Once a region of interest has been found, the scan in the conventional SPM software is paused and the FAST module operation can be started. The high-speed ADCs are used to acquire the tunneling current, which is, in this case, the response signal of interest. Indeed, by acquiring the logarithm of the tunneling current, a good approximation of the actual topography can be obtained. Any phase shift between the generated scan signals and the measured tunneling current response is manually corrected through a pixel shift control of the visualized data in the GUI. When using a high scan speed, the scanner is brought close to resonance conditions and the amplitude of its response is no longer linear with the applied voltage. A proper calibration procedure is therefore required to know the correct image size, which, again, needs to be performed just once for a specific set of scanning parameters. Besides the conventional *a posteriori* recognition of features of known periodicity on the acquired image, the calibration can be conveniently accomplished *a priori* by simply changing the scanner position by a known quantity through the conventional scanning software while imaging a certain feature with the FAST module.

As mentioned above, one of the strengths of the FAST module is to enable time-resolved SPM studies, without stringent requirements on the microscope/scanner design. This allows, for example, its implementation in experimental setups already optimized for *in-situ* studies, capable then to work at high temperatures and pressure, as well as in electrochemical environments. In this context, the new FastSTM has been already successfully exploited for the investigation of a variety of dynamical processes occurring at surfaces, such as catalytic reactions [21, 22], assembly of supramolecular networks [23] and diffusion of supported metal clusters [24]. These recent results demonstrate the actual possibility to study surface processes at the atomic scale in real-time, providing new insight and understanding on the dynamics occurring in the millisecond time scale.

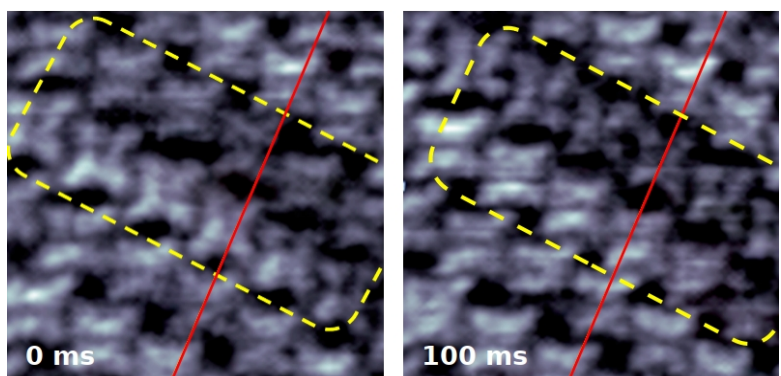

Figure 2 - Collective motion of two rows of tetra-aminophenyl porphyrin (TAPP) molecules on Au(111) (encircled in yellow) in consecutive FastSTM frames. Red lines serve as guides to better visualize the shift (110 × 110 pixels, 1.1 kHz fast scan, 10 fps, 8 × 8 nm$^2$, -1.0 V, 0.15 nA, 300 K).

The example reported in Figure 2 shows two frames of a FastSTM movie acquired on a self-assembled monolayer of tetra-aminophenyl porphyrin molecules on Au(111). The consecutive frames show the collective dynamics of two rows of molecules at room temperature on a time scale of 100 ms. This example shows two important features of the FAST module. First of all, it is possible to image organic molecules that can be weakly bonded to a surface, with limited scanning-induced structural perturbations, thanks to the ability to image also at relatively low tunneling currents. Secondly, it can be clearly seen that the imaging resolution is comparable with that typically achievable in the slow scanning mode, allowing to observe also the intra-molecular structure of the molecules.

## 5. FastAFM

AFMs, where the deflection of a cantilever beam due to the attractive/repulsive interactions between a sharp tip and a surface is detected, can be used in different environments (vacuum, air, gas and liquid) for a plethora of materials with differentiated physical properties (stiff, soft, conductive, insulating, etc.), thus covering a vast range of scientific and technical applications in materials sciences, microelectronics, biology and metrology, among others [25]. In the vast majority of cases, atomic resolution is not a requirement in the multidisciplinary AFM community. Instead, the focus lies on the meso-scale, with nanometer-size features (lateral dimensions typically above 10 nm) and sampling areas usually above 1 × 1 µm$^2$. Acquisition of images with sufficient resolution is typically rather slow: 256 lines scanned at 1 Hz (1 sec per line) result in more than 4 minutes per image. Sometimes, the surface can irreversibly evolve during image acquisition due to the interaction with the tip especially in the case of soft samples. Thus, there is a real need to reduce the acquisition time for practical purposes and also to open the possibility to study dynamical processes in real time and *operando*.

Recent progresses in high-speed AFMs have been achieved since the first report in 1991 [10] thanks to the development of ultra-short cantilevers [11], which exhibit resonance frequencies of 1 MHz and above, higher than those typically used in dynamic mode, and to the development of dedicated high-speed scanners, including counterbalanced [12], flexure-based [13] and MEMS-based scanners [14], to mention but a few, together with optimized fast amplitude detectors and feedback electronics [15]. Commercial AFMs designed to operate in fast scan mode exhibit frame rates typically in the 10 frames per second (fps) range, as claimed by the manufacturers.

We applied the FAST module to a commercial AFM instrument, Keysight Technologies' 5500 (formerly Agilent Instruments [PicoPlus] 5500), normally working at 0.004 fps for a typical image with good resolution, to operate it in fast scan mode. Frame rates above 8 fps in static (contact) mode and above 1 fps in dynamic (amplitude modulation) mode were consistently achieved on silicon and PS-PMMA block copolymer reference samples, with a pitch of about 38 nm [26].

Compared to state-of-the-art instruments, our test system exhibits limited performance in terms of (i) photodetector bandwidth (~ 910 kHz), which hinders the use of cantilevers with resonance frequencies above 1 MHz, and (ii) the use of a laser spot that is too large for ultra-short cantilevers. In spite of these technical limitations, it can be consistently operated in the fast scan mode, as shown below.

In our experimental setup, a Keysight N9447A breakout box is used to exchange scan signals between the FAST module and the host AFM system using standard 50 Ω BNC coaxial connectors. The AFM high voltage positioning signals are branched out and fed into the FAST module, in which the fast scan signals are modulated onto the positioning "carrier" signal and then routed back to the AFM system in order to be applied to the piezoelectric motor position, just as the carrier signal alone would be. The measurement variable used for the fast system depends on the chosen mode of AFM operation: (raw) deflection for static/contact AFM or amplitude and/or phase signal for dynamic/tapping AFM, which are accessible via 50 Ω BNC connectors located either on the Head Electronics Box or AC controller box of the AFM system. In the setup arrangement one computer is dedicated to run the Keysight AFM software (PicoView 1.20.3) thus providing a complete, independent conventional AFM system. A second PC runs the FAST control and acquisition software.

We have used the Keysight N9524A multipurpose scanner which has nominal lateral and vertical scan ranges of 90 µm and 7 µm, respectively, for the available drive voltages of ±215 V per axis that the Keysight controller can deliver. The piezoelectric actuator is mounted in a frame, which holds the optical system (laser and photodetector) and a nose cone with a window conceived to work in liquid environment and the cantilever holder. The spectral response of the scanner to frequencies up to 1 kHz exhibits a resonance at about 420 Hz.

The sample slope compensation is a key factor to consider for AFM since the target scan widths are noticeably larger than those used in STM under atomic resolution conditions. If such a slope cannot be efficiently compensated and for samples with topography features in the low nanometer range, the feedback cannot be completely disabled, with a resulting penalty in terms of acquisition time. The FAST system does not provide a feedback loop for vertical position control of the scanned probe. Just like in the FastSTM setup described in the previous section, the FastAFM measurements shown below were performed in quasi-constant height mode, in which the influence of the feedback loop of the original AFM system is minimized to make its response much slower than the probe movement along the fast scan direction ($x$), while simultaneously being fast enough to compensate for sample tilt along the slow scan direction ($y$) and sample/scanner drift in vertical directions. Obviously, the optimal feedback configuration varies with the chosen frame rate and physical scan range.

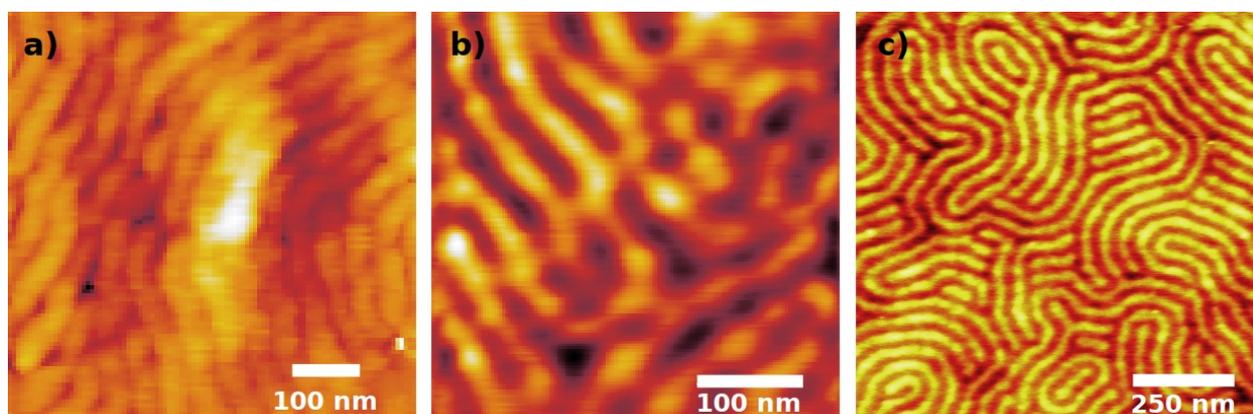

Figure 3 - Frames from FastAFM movies of a fingerprint-like silicon surface acquired in contact mode (a) and a fingerprint-like PS-PMMA block copolymer thin film acquired in tapping mode (b). The estimated scanned area is (a) 600 × 600 nm$^2$ and (b) 385 × 385 nm$^2$, respectively, and the line

frequency of the fast scan axis is (a) 800 Hz and (b) 150 Hz, respectively. (c) Topography image acquired in standard tapping mode of the fingerprint-like PS-PMMA block copolymer thin film.

Figure 3(a) shows a frame from a FastAFM movie of the patterned silicon sample in air acquired in contact mode at 8 fps and a resolution of 100 × 100 pixels. In this configuration, the line frequency is equivalent to 800 Hz and the covered area corresponds to about 600 × 600 nm$^2$. The image has been acquired using a PPP-CONTSCR cantilever (Nanosensors) with a nominal force constant of 0.2 N/m. Above 8 fps, the images become unstable although the fingerprint structure can still be identified. Figure 3(b) shows a frame of a FastAFM movie obtained from the phase signal when scanning the PS-PMMA block copolymer sample in tapping mode using a qp-fast cantilever (Nanosensors) with a resonance frequency of 871 kHz and a nominal force constant of 80 N/m. The explored area corresponds to about 385 × 385 nm$^2$ and was scanned at a frame rate of 1 Hz with a lateral resolution of 150 × 150 pixels which corresponds to a line frequency of 150 Hz. Because of the variation of sensitivity of piezoelectric actuators with scanning rates, the periodicity of the fingerprint structures was used for lateral dimension calibration by comparing images acquired in fast scan mode with those obtained for the same surface in the corresponding standard AFM mode (see Figure 3(c) acquired at 1 Hz line frequency). The scans on the patterned Si structures show no morphological changes during acquisition (indicating a stable tip and surface). The integral (I) and proportional (P) gains of the feedback were I = 0.5 and P = 0 in contact mode and I = 0.008 and P = 0 in tapping mode, respectively. In tapping mode, the trade-off of sampling frequency versus resonant frequency is crucial so that the number of oscillations per pixel is satisfactory. In our case, with a cantilever resonance frequency of 871 kHz, a line frequency of 150 Hz and 150 pixels, the estimated number of oscillations per pixel is 20.

## 6. Atom tracking

While FastSTM/AFM measurements usually require elevated signal levels (tunneling currents of several hundreds of pA) due to higher noise levels at increased bandwidth of the preamplifier, a high time resolution can be obtained as well with low signal levels when switching from the fast scan mode to the atom tracking mode, where the tip simply follows the lateral and vertical movement of a local extremum (protrusion or depression) on which it is centered by a lateral, two-dimensional feedback. This technique, proposed by Pohl and Möller [17], has been implemented for the first time by Swartzentruber for the study of adatom diffusion [16]. It is currently included in advanced scanning probe controllers for accurate drift control [27-29]. In this technique, circular dithering of the tip with a typical frequency below 1 kHz and a dithering radius of the order of 1 nm and below, combined with a phase sensitive detection in *x* and *y* direction, allows for determining the respective error signals that can be directly applied as input for a lateral feedback via a proportional-integral-derivative (PID) controller [30]. In this way, the dynamics of single species can be followed with time resolution down to the ms and sub-Å lateral resolution, while being able to simultaneously measure the topographic *z*-channel and, optionally, a second channel such as the current/conductance. The low dithering frequencies and thus the low required bandwidth in the kHz range allow for measurements with currents down to the pA. Atom tracking can thus deliver precious additional information, as a complementary, low-signal technique to FastSTM/AFM measurements. The immediate access to displacement rates allows for systematic investigation of the probe influence on the investigated particle dynamics. Bias and tip current dependencies, for example, help to judge the effects of electric field and direct chemical interactions [31]. Its seamless integration into the FAST module is therefore of fundamental interest.

To that purpose, a separate LabVIEW control program for atom tracking has been developed that uses the input/output connections in a slightly different signal mapping than FastSTM/AFM, as described in Figure 1: while the two fast analog output channels are used for generating the dither signal, the two fast analog input channels detect the topographic z-channel and current/conductance (in the following we refer just to the STM example, even if the technique can be equivalently used for AFM). Two slow analog output channels are used to apply the offset voltages to the *x* and *y* piezoelectric actuators quadrants that are delivered by the PID controller. Similar to the FastSTM/AFM program, the data are acquired with the same FPGA code, in blocks of 1 ms duration that are then analyzed by a digital lock-in. Typically the resulting error signals are subsequently averaged with a moving average over 10 values, leading to an overall time resolution of 10 ms. This time constant could, in principle, be improved by using an analog lock-in, as long as the interfacing is not conflicting with the user-friendliness.

User-friendliness is, however, extremely important for efficient measurements. This concerns, first of all, the interconnection of the FAST module with the slow, standard STM controller via transistor-transistor logic (TTL) control signals: when repositioning the tip over a particular feature with the slow controller, the dither movement and, after a certain delay of about 100 ms, the lateral feedback start automatically, making it possible to jump on fast moving objects to track them. At the end, after the atom tracking measurement, the last applied *x* and *y* offset voltages continue to be output by the FAST module, while the lateral feedback is stopped, in order to keep the observed feature well-centered for a subsequent imaging by the slow standard STM controller when it comes to overlay diffusion trace and local topography and to control the actual environment of the diffusing feature.

User-friendliness concerns, secondly, the proper setup of the control parameters. The frequencies at which the dithering movement is excited are close to bending eigenfrequencies of the piezoelectric scanner and might hence distort the movement from an accurate circular trace. Therefore, a proper regulation of the excitation phase (between *x* and *y* movement), the excitation amplitudes and of the detection phase is necessary. When properly regulated, the error signal in *x* over a hemispherical protrusion will show a vertical transition between positive and negative errors and that in *y* a horizontal one. An example is given in Figure 4(a) that shows a $Pd_{12}$ cluster deposited on a graphene moiré film grown on Rh(111). This surface of a periodicity of 3.2 nm is periodically wettable, with two sticky sides in each elementary cell, where the clusters bind preferentially [32, 33].

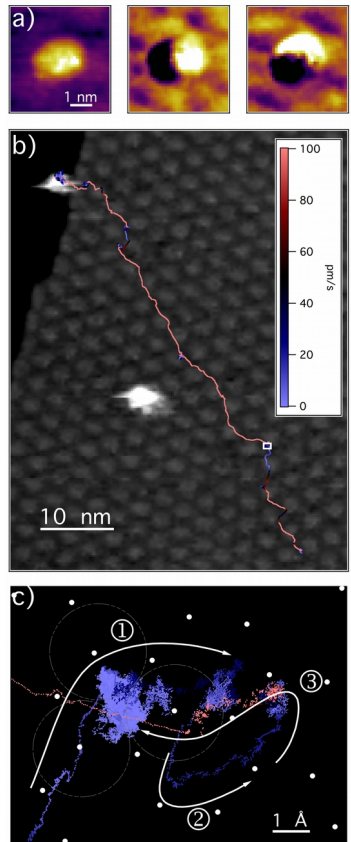

Figure 4 - Atom tracking experiment performed with the FAST module: $Pd_{12}$ cluster diffusing on a periodically wettable moiré lattice of graphene on Rh(111). Imaging conditions: +1 V, 1 pA. (a) Images of cluster topography (left) and corresponding error signal in the *x*- (middle) and *y*-direction (right) upon dithering the tip (800 Hz, 1 nm radius). (b) Overlay of the velocity-encoded atom tracking trace on an STM image obtained after the diffusion ceased on the lower terrace side of a step edge. The cluster could be traced over 30 min and 100 nm distance while maintaining a temporal resolution of 10 ms and a lateral one below 1 Å. (c) Enlarged detail (area indicated by white rectangle in lower right half of (b), arrows indicate the overall movement in the numbered sequence).

Displayed are the topographic *z*-channel and the *x* and the *y* error signals of a well-corrected tip movement, such as they can be measured within the atom tracking program. These images have been taken with a specific imaging routine that can be called and that performs a slow lateral scan, with which the tip and the setting qualities are tested prior to the tracking measurement. Since the cluster in this example is two atomic layers high (*i.e.* itself a sharp tip), the measuring tip has to be of adequate cylindrical symmetry. The dithering movement of 1 nm diameter is hardly recognizable in the topographic *z*-channel due to overlapping tip convolution effects.

In a last step, the feedback parameters have to be optimized. Best results have been obtained with a PI controller, where the integral time constant was about 10 ms, on the order of the lock-in time constant. As a test for the proper functioning and as a means for calibration, defined *x* and *y* offset steps of some nm width can be induced by the slow controller. They have to be mapped adequately by the atom tracking program - in particular, an apparent curvature of the traces indicates erroneous phase settings.

With a properly adjusted tracking setup, diffusion traces such as in Figure 4(b) can be measured. In this example, the cluster diffuses over distances of 100 nm in about 30 minutes, before being caught on the lower terrace side of a step edge, where the cluster movement gets blocked, as indicated by the image taken after the tracking measurement. The diffusion trace is drift-corrected by subtracting a smooth drift vector gained from an analysis of the difference vectors between adjacent $x$ and $y$ values. The drift trace indicates that the diffusing cluster avoids the protruding, non-wettable areas of the moiré structure. Furthermore, the velocity-encoding shows that the diffusion occurs in a very discontinuous way (with velocities smoothed with a moving average over 1 s), pointing to anomalous diffusion (superdiffusion): once the cluster starts to move, it diffuses notably, far more than one elementary cell of the moiré structure, having difficulties to slow down again. A detail of such a rare dissipative event is shown in Figure 4(c) (enlargement of the white rectangle in Figure 4(b)). The movement of the cluster apex that is indicated by the arrows can be resolved with sub-Å lateral resolution and 10 ms temporal resolution. As a guide to the eye, the periodicity of the carbon lattice of the graphene structure is indicated by white dots, while the Pd and Rh atom size is schematically drawn by the three circles - the cluster diameter (footprint of seven atoms, as obtained by calculations [32]) is therefore larger than the displayed image size.

Summing up, the FAST module allows for a seamless integration of the atom tracking functionality with a maximum content of pre-configured routines that increase the user-friendliness in order to guarantee a successful outcome of demanding measurements. The complementarity of atom tracking and FastSTM/AFM measurements opens up the full panorama of SPM investigation techniques to capture particle dynamics: while tracking unveils the dynamics of individual particles, fast imaging unveils the dynamics of several particles and hence of the statistical differences between particles, e.g. due to surface heterogeneities or isomer distributions. The low signal levels required for tracking open up measurement regimes where the probe influence on the investigated particle dynamics is minimal.

## 7. Portability and transferability of the module

The FAST module can in principle upgrade any scanning probe microscope to video-rate imaging, the only requirement being the possibility to breakout the high voltage signals to the SPM head. The module has been initially developed and thoroughly tested on an Omicron VT-STM operated by SCALA electronics, on which a frame rate up to 100 Hz at a resolution of 100 × 100 pixels can be routinely reached. The portability of the FAST module has been successfully proven also on a similar Omicron VT-instrument including AFM optics and on an Omicron LT-STM, both driven by Matrix electronics. Atomic resolution has been obtained up to a frame rate of 60 Hz (using the LT-STM at 77 K) and temperatures up to 780 K (using the VT-STM/AFM at a frame rate of 12 Hz). Preliminary tests have been conducted also on a SPECS Aarhus VT-STM and an STM driven by Nanotec electronics at RT reaching frame rates up to 40 Hz. In all cases the integration of the FAST module proved to be an easy and rapid procedure, with the reported performance reached in a matter of hours (see Figure 5).

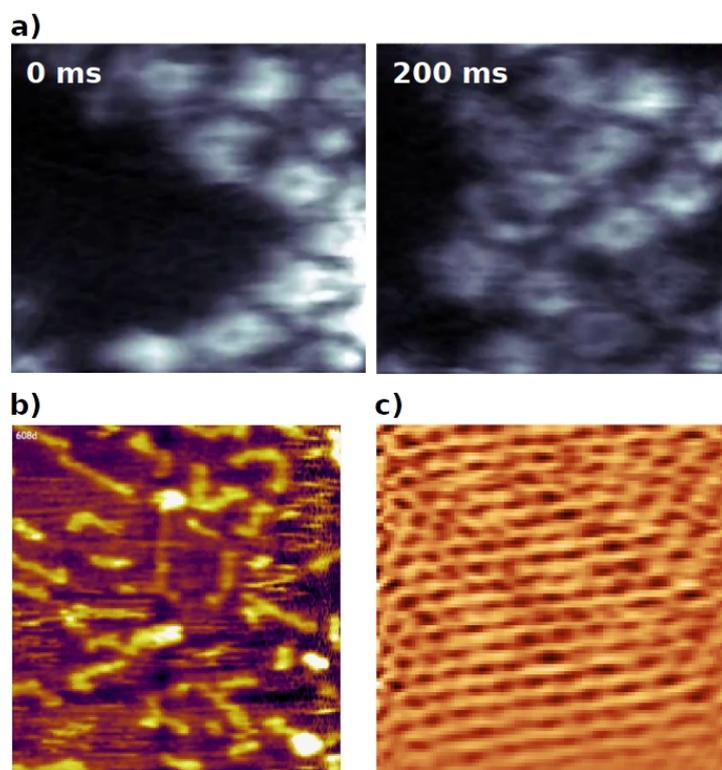

Figure 5 - Frames from FastSTM movies acquired with the FAST module coupled to different instruments. (a) Self-assembly of tetra-aminophenyl porphyrins on Au(111) at 77 K taken with an Omicron LT-STM at 61 fps. (b) Polymerization of ethene taken with an Omicron VT-STM/AFM, 4 fps, 14 × 14 nm$^2$ [34]. (c) Graphite at RT taken with an STM driven by Nanotec electronics at 40 fps.

## 8. Data conditioning and analysis

The FastSTM and FastAFM movies are recorded as a one-dimensional (1D) raw data stream in an HDF5 file, in order to allow for full control on reconstruction and post-processing operations. To correctly transform the 1D data into a 3D movie stack, we use a specially written Python 3 software package, *pyfast*, which will be described in detail elsewhere. *Pyfast* assigns a location to each original pixel in order to map the data stream onto an equally spaced square mesh. To obtain the specific coordinates, the movement of the tip is described by a combination of a sinusoidal oscillation in the fast scan direction, a simultaneous triangular movement in the slow scan direction that additionally includes a piezoelectric creep at the turnaround point, and a continuous thermal drift. The mapping then occurs via linear or cubic interpolation taking into account the lateral positions frame-by-frame. In this way, movies can be created that utilize 100% of the recorded data points by interlacing - *i.e.* weaving into each other - the forward and backward lines into a single frame with twice the number of data points. Equally, the frame rate is maximized by accurately correcting the distortion between upward and downward frames.

In addition to the reconstruction of the image stack, *pyfast* includes some 1D and 2D filters which remove noise and achieve plane correction by filtering the frequencies of the fast and slow scan directions. Care is taken not to reduce the time resolution of the movie when filtering. In a final step, the movie can be exported into an H.246/MP4 file with variable compression, or alternatively into images of single frames in common image file formats.

Compared to the FastSTM/AFM evaluations, the evaluation of the atom tracking data is quite simple. The trajectory is recorded as a time-sequence of *x* and *y* coordinates and requires exclusively a drift subtraction for the *x*, *y* and *z* data. This drift correction is performed by subtracting a continuous directed trend that can easily be discerned from random walk movements by appropriate smoothing of difference values and thresholding.

## 9. Conclusions and outlook

We have presented the new version of FAST, an electronics module capable of accelerating commercial scanning probe microscopes to video frame rates and beyond. The new module, as its prototype, transparently uses LabVIEW components and a summation board which are inserted between the standard control electronics and STM or AFM microscopes, but now the LabVIEW software package provides a user friendly, simple yet comprehensive interface to measure video rate movies at a location that is pre-defined in the standard control software. The measured movies are processed and converted into common video and image formats by a custom Python software package. We have demonstrated the versatility of the FAST module, showing data recorded on different commercial STM instruments. Furthermore, the same module has been successfully used to drive a commercial AFM, demonstrating the potential interest of the device for an extremely large user community.

We achieved frame rates up to 100 fps with FastSTM and 8 fps with FastAFM. In the case of FastSTM the highest achievable speed is limited mainly by the cutoff frequency of the preamplifier, thus virtually allowing to achieve higher frame rates.

In addition, the module can be used to track the motion of particles with the tip by atom tracking, *i.e.* dithering the tip in *x* and *y* to keep it on top of a protrusion or depression, thereby increasing the time resolution even further. This versatile and simple approach paves the way for dynamical studies with standard SPM instruments down to the millisecond time scale for the general user community. This can be of real interest to many laboratories around the world which use these instruments on a daily basis, with the added advantage that the upgrade to the fast scan mode can be achieved without any modification of the existing hardware and of the control electronics.

## Acknowledgements

The FAST project has received funding from the EU-H2020 research and innovation programme under grant agreement No 654360 NFFA-Europe. The FAST module has been developed to be at public disposal and is hence also commercially available [35]. The ICN2 is funded by the CERCA program/ Generalitat de Catalunya. The ICN2 is supported by the Severo Ochoa program of MINECO (Grant SEV-2017-0706). We acknowledge S. Gottlieb, M. Fernández-Regúlez and F. Pérez-Murano for providing us with the silicon and block copolymer samples. Work at TUM was supported by the Deutsche Forschungsgemeinschaft (Research grants ES 349/1-2 and HE 3454/18-2). BAJL acknowledges a Research Fellowship from the Alexander von Humboldt Foundation and a Marie Skłodowska-Curie Individual Fellowship under grant ClusterDynamics (no. 703972) from the EU Horizon 2020 Programme.